\documentstyle[12pt]{article}
\input epsf
\begin{document}

\newcommand{\be}{\begin{equation}}
\newcommand{\ee}{\end{equation}}
\def\bq{\begin{eqnarray}}
\def\eq{\end{eqnarray}}

\begin{center}
\large{{\bf Einstein-de Sitter model re-examined for the newly discovered SNe Ia}}
\end{center}

\begin{center}
R. G. Vishwakarma,\footnote{E-mail: rvishwa@mate.reduaz.mx}\\
\emph{Unidad Acad$\acute{e}$mica de Matem$\acute{a}$ticas\\
 Universidad Aut$\acute{o}$noma de Zacatecas\\
 C.P. 98068, Zacatecas, ZAC.\\
 Mexico}\\
\end{center}

\bigskip
\begin{abstract}\noindent
Consistency of Einstein-de Sitter model with the recently observed
SNe Ia by the Hubble Space Telescope is examined. The model shows a reasonable
fit to the observation, if one takes into account the extinction of SNe light
by the intergalactic metallic dust ejected from the SNe explosions. Though the 
fit to the new data is worsened considerably compared with the earlier data, it
can still be regarded acceptable. We should wait for more accurate observations
at higher redshifts (as expected from the coming space missions such as SNAP 
and JWST) in order to rule out a model, which seems to explain all
the other existing observations well (some even better than the favoured 
$\Lambda$CDM model), is consistent with beautiful
theoretical ideas like inflation and cold dark matter, and is not as 
speculative as the models of dark energy.

\medskip
\noindent
{\it Subject heading:} cosmology: theory, whisker dust, SNe Ia.

\end{abstract}

\noindent
{\bf 1. INTRODUCTION}

\noindent
The high redshift supernovae (SNe) Ia explosions look fainter than they are 
expected in the
Einstein-de Sitter (EdS) model (flat cold dark matter model with 
$\Omega_{\rm total}=1$), which used to be the favoured
model before these observations were made a few years ago. 
The most popular explanation of this observed faintness is given
by invoking \emph{`dark energy'}, a homogeneously distributed energy 
component with negative pressure.
This happens because the metric distance of an object, out to any redshift, 
can be increased by incorporating a \emph{`fluid'} with negative
pressure in Einstein's equations.
The simplest and the most favoured candidate of dark energy is Einstein's 
cosmological constant $\Lambda$.

Alternative mechanisms have also been suggested in order to explain the 
observed faintness of the distant SNe Ia, without invoking cosmic acceleration.
One of such attempts considers photon-axion oscillations in the
intergalactic magnetic field as a 
way of rendering supernovae dimmer without invoking dark energy 
(Csaki, et al 2002; 2004).

Another alternative way to explain the faintness of the high redshift SNe Ia
is to consider the absorption of light by metallic dust 
ejected from the SNe explosions$-$an issue
which is generally ignored while discussing $m$-$z$ relation
for SNe Ia. 
It was shown earlier (Vishwakarma, 2003) that if this effect is taken 
into account, the EdS model explains the SNe Ia data 
from Perlmutter, et al (1999) (together with SN 1997ff at $z\approx1.7$) 
successfully. 
Since then, many SNe Ia at higher redshifts have been discovered  with
the Hubble Space Telescope (HST) (Riess, et al 2004).
Extending our earlier work further, we examine in this paper how well or badly 
the EdS model fits the new data which include 7 highest redshift SNe Ia known
so far, all at $z>1.25$.

\bigskip
\noindent
{\bf 2. EXTINCTION BY METALLIC DUST}

\noindent
Chitre and Narlikar (1976) were the first to discuss the role of
intergalactic dust in the $m$-$z$ relation, which was however largely ignored 
that time.
It is, however, generally accepted now that the metallic vapours are ejected 
from the SNe
explosions which are subsequently pushed out of the galaxy through pressure
of shock waves (Hoyle \& Wickramasinghe, 1988; Narlikar et al, 1997).
Experiments have shown that metallic vapours on cooling, condense into
elongated whiskers of $\approx$ $0.5-1$ mm length and $\approx$$10^{-6}$ cm 
cross-sectional radius (Donn \& Sears, 1963; Nabarro \& Jackson, 1958).
It has been shown, by considering standard big bang and alternative
cosmologies, that
this kind of dust extinguishes radiation travelling over long distances
(Aguire, 1999; Banerjee, et al 2000; Narlikar, et al 2002; Vishwakarma, 2002; 
2003). 
The dust presumably re-radiates the absorbed radiation at another 
wavelength. For example, it degrades a radiation of high-frequency 
into the far infrared. The 
imergent radiation is then redshifted into the millimeter waveband due to
the expansion of the universe. However, this is not the subject of the 
present paper. 
The density of the dust can be estimated
along the lines of Hoyle et al (2000): If the metallic whisker production is
taken as 0.1 $M_\odot$ per SN and if the SN production rate is taken as
1 per 30 years per galaxy, the total production per galaxy (of spatial density
$\approx$ 1 per $10^{75}$ cm$^3$) in $10^{10}$ years is $\approx 2/3\times 
10^{41}$ g. The expected whisker density, hence, becomes 
$2/3\times 10^{41}\times 10^{-75}\approx 10^{-34}$ g cm$^{-3}$.
We shall see later that this value is in striking agreement with the
best-fitting value coming from the SNe Ia data.

In the following we calculate the mean extinction of the SNe light from the 
whisker dust.  Suppose that the light we observe today was emitted at the 
epoch of redshift $z$ from a SN of luminosity $L$.
In traversing a distance  $\rm d$$\ell$ through the 
intergalactic medium, the light looses its brightness by an amount $\rm d$$L$ 
due to the absorption by the whisker grains of mean density $\rho_{\rm g}$. 
This is given by
\be
{\rm d} L=-\kappa ~\rho_{\rm g} ~L ~\rm d \ell,\label{eq:delL}
\ee
where $\kappa$ is the mass absorption coefficient, which is effectively
constant over a wide range of wavelengths and is of the order $10^5$
cm$^2$ g$^{-1}$ (Wickramasinghe \& Wallis, 1996). Hence the total loss
in the luminosity $\Delta L$ of SN light, 
in traversing a proper distance $\ell(z)$, can be obtained by integrating 
equation (\ref{eq:delL}):
\be
\Delta L(z)=\exp\left[-\kappa \int_0^{\ell(z)}\rho_{\rm g} ~\rm d 
\ell\right].
\label{eq:DelL}
\ee
For the Robertson-Walker (RW) metric, this reduces to
\begin{equation}
\Delta L(z)=\exp\left[\kappa ~\rho_{\rm g0}\int_z^0
(1+z')^2\frac{{\rm d}z'}{H(z')}\right],
\end{equation}
where $\rho_{\rm go}$ is the whisker grain density at the present epoch and
$\rho_{\rm g}~S^3=\rho_{\rm g0}~S_0^3=$ constant, $S(t)$ being the scale 
factor of the RW metric. Here and henceforth we have considered the speed
of light $c=1$. The corresponding increase in the
magnitude $\Delta m$ ($=-2.5 \log \Delta L$) is thus given by
\begin{equation}
\Delta m(z)=1.0857\times ~\kappa ~\rho_{\rm g0}\int_0^z
(1+z')^2\frac{{\rm d}z'}{H(z')}.\label{eq:deltam}
\end{equation}
The net magnitude is then given by
\begin{equation}
 m^{\rm net}(z)=m(z) + \Delta m(z),\label{eq:mnet}
\end{equation}
where the first term on the right corresponds to the usual apparent magnitude 
of the object resulting from the cosmological evolution:
\be
m (z) =5 \log[H_0 ~d_{\rm L}(z)] + {\cal M},\label{eq:mageq}
\ee
with the luminosity distance $d_{\rm L}$ given by
\be
d_{\rm L}(z) = (1 + z) \int_0^z \, \frac{{\rm d} z'}{H(z')},
\label{eq:distL}
\ee
for the $k=0$ case of the RW metric. The constant ${\cal M}$ appearing in 
equation (\ref{eq:mageq}) is given by 
${\cal M} \equiv M - 5 \,\log H_0 + constant$, where $M$ is the absolute 
magnitude of the SNe. For the EdS model ($\Omega_{\rm m0}=1$, 
$\Omega_{\Lambda0}=0$), equation (\ref{eq:mnet}) reduces to
\be
m^{\rm net}(z)= 5 \log[2\{(1+z)-(1+z)^{1/2}\}] 
+ 0.7238\times ~\kappa ~\rho_{\rm g0}~H_0^{-1}[(1+z)^{3/2}-1]  
+ {\cal M}.\label{eq:mEdS}
\ee

\bigskip
\noindent
{\bf 3. DATA FITTING}

\noindent
We consider the data recently published by Riess et al (2004) which, in 
addition to
having previously observed SNe, also include 16 newly discovered SNe Ia, 
6 of them being among the 7 highest redshift SNe Ia known, all at 
redshift $>1.25$. We particularly focus on their `gold sample' of 157 SNe Ia
which is believed to have a `high confidence' quality of the spectroscopic and
photometric record for individual supernovae. We note that the data points of
this sample are given in terms of distance modulus 
$\mu_o=m^{\rm net}-M=5\log d_{\rm L}+constant$. However, the zero-point 
absolute magnitude 
were set arbitrarily for this sample. Therefore, while fitting
the data, we can compare the observed $\mu_o$ with our predicted $m^{\rm net}$
given by equation (\ref{eq:mEdS}) and compute $\chi^2$ from
\be
\chi^2 = \sum_{i = 1}^{157} \,\left[ \frac{m^{\rm net}(z_i;
~\kappa \rho_{\rm g0}H_0^{-1}, ~{\cal M}) - \mu_{o, i}}
{\sigma_{\mu_{o, i}}}\right]^2.\label{eq:chi}
\ee
The constant ${\cal M}$ thus plays the role of the normalization constant.
The quantity $\sigma_{\mu_{o, i}}$ is the uncertainty in the distance
modulus $\mu_{o, i}$ of the $i$-th SN.

There are only two  free parameters in this model which are 
to be estimated from the data: $\kappa \rho_{\rm g0}H_0^{-1}$ and ${\cal M}$.
The parameter $\kappa \rho_{\rm g0}H_0^{-1}$, which is 
dimensionless, is of the order of unity if one considers $\kappa$ of the order
$10^5$ cm$^2$ g$^{-1}$, $\rho_{\rm g0}$ of the order $10^{-34}$ g cm$^{-3}$ and
$H_0\sim$ 70 km s$^{-1}$ Mpc$^{-1}$.
 However, we have kept it as a free parameter to be estimated from the data.

By varying the free parameters of the model, we find that the minimum $\chi^2$ 
is obtained for ${\cal M}=43.41$ and $\kappa \rho_{\rm g0}H_0^{-1}=4.77$
(where $\kappa$, $\rho_{\rm g0}$ and $H_0$ have been measured
in units of $10^5$ cm$^2$ g$^{-1}$, $10^{-34}$ g cm$^{-3}$ and
100 km s$^{-1}$ Mpc$^{-1}$ respectively). These give a value $\chi^2=200.99$
at 155 degrees of freedom (dof), that is, $\chi^2$/dof $= 1.29$, which 
represents a reasonable fit. In the absence of the extinction from the 
whisker dust, i.e. for
$\rho_{\rm g0}=0$, one obtains a too high $\chi^2$/dof $= 324.70/156=2.08$,
with ${\cal M}=43.58$.

Though there is not a clearly defined value of $\chi^2$/dof for an acceptable
fit, a \emph{`rule of thumb'} for a \emph{moderately} good fit is that $\chi^2$
should be roughly equal to the number of dof. A more quantitative measure
for the \emph{goodness-of-fit} is given by the $\chi^2$-\emph{probability}
which is very often met with in the literature and its compliment is usually
known as the \emph{significance level} (should not be confused with the confidence
regions). If the fitted model
provides a typical value of $\chi^2$ as $x$ at $n$ dof, this probability is
given by
\be
Q(x, n)=\frac{1}{\Gamma (n/2)}\int_{x/2}^\infty e^{-u}u^{n/2-1} {\rm d}u.
\ee
Roughly speaking, it measures the probability that \emph{the model does
describe the data and any discrepancies are mere fluctuations which could have
arisen by chance}. To be more precise, $Q(x, n)$ gives the probability that a model
which does fit the data at $n$ dof, would give a value of $\chi^2$ as large
or larger than $x$. If $Q$ is very small, the apparent discrepancies are
unlikely to be chance fluctuations and the model is ruled out. It may however
be noted that the $\chi^2$-probability strictly holds only when the models are
linear in their parameters and the measurement errors are normally distributed.
It is though common, and usually not too wrong, to assume that the 
$\chi^2$- distribution holds even for 
models which are not strictly linear in their parameters, and for this reason,
the models with a probability as low as $Q>0.001$ are usually deemed 
acceptable (Press et al, 1986). 
Models with vastly smaller values of $Q$, say, $10^{-18}$ are rejected.

The probability $Q$ for the best-fitting EdS model is obtained as 0.008, 
which is though very small, but acceptable. In the absence of the whisker dust,
this probability reduces to $1.5\times10^{-12}$, which is too small to be 
accepted.
In order to compare these results with those in the $\Lambda$CDM
cosmology, we note that for a constant
$\Lambda$, the best-fitting models are obtained as

\medskip
\noindent
Global best-fitting model: $\Omega_{\rm m0}=0.46$, $\Omega_{\Lambda0}=0.98$, 
${\cal M}=43.32$, with $\chi^2$/dof $= 175.04/154=1.14$, $Q=0.118$;\\
Best-fitting flat model: $\Omega_{\rm m0}=1-\Omega_{\Lambda0}=0.31$, 
${\cal M}=43.34$, with $\chi^2$/dof $= 177.07/155=1.14$, $Q=0.109$;\\
Concordance model ($\Omega_{\rm m0}=1-\Omega_{\Lambda0}=0.27$): 
$\chi^2$/dof $= 178.17/155=1.15$ with $Q=0.098$.

\medskip
\noindent
We note that the fit to the EdS model, even with the whisker dust, is 
considerably worse than those in the $\Lambda$CDM models. 
However we also note that the new data has a general tendency to worsen the 
fit to any model. For example, the earlier data with 57 points (Vishwakarma, 
2003) gave better fits to the models mentioned above:

\medskip
\noindent
EdS model (with whisker dust): $\chi^2$/dof $ = 68.97/55 = 1.25$, $Q=0.098$;\\
Best-fitting $\Lambda$CDM: $\Omega_{\rm m0}=0.64$, $\Omega_{\Lambda0}=1.20$ 
with $\chi^2$/dof $=57.78/54=1.07$, $Q=0.337$;\\
Best-fitting flat $\Lambda$CDM: $\Omega_{\rm m0}=1-\Omega_{\Lambda0}=0.32$ with
$\chi^2$/dof = 59.67/55 = 1.08, $Q=0.31$.

\medskip
\noindent
We note that the new data have worsened the fit to the EdS model to a greater
extent compared to the $\Lambda$CDM model, however, considering the successes
of the EdS model, we should be tolerant of the low probability associated 
with it, until it goes down to some value like the one obtained without the 
whisker dust. The reasons are many:

\medskip
\noindent
(i) The EdS model is fully 
consistent with the recent CMB observations made by the WMAP. In fact, there is
a degeneracy in the $\Omega_{\rm m0}-\Omega_{\Lambda0}$ plane along a line
$\Omega_{\rm m0}+\Omega_{\Lambda0}\approx 1$ and a wide range of $\Omega_{\rm m0}$ is consistent with the observations (Vishwakarma, 2003). For example, with
$\Omega_{\rm b0}=0.1$ and $H_0=55$ km s$^{-1}$ Mpc$^{-1}$, this model
yields the positions of the first three acoustic peaks at the multipole values
$\ell_{\rm peak_1}=220.4$,
$\ell_{\rm peak_2}=521.4$,
$\ell_{\rm peak_3}=784.9$, which are in good agreement with WMAP. Additionally
a lower value of $H_0$ is favourable for this model to have sufficient age. 
For example, $H_0$ should be $\leq 54$ km s$^{-1}$ Mpc$^{-1}$ for EdS model
to have an age of the universe $\geq 12$ Gyr, so that the age of the
oldest objects detected so far, e.g., the globular clusters of age 
$t_{\rm GC}=12.5\pm 1.2$ Gyr (Cayrel et al. 2001; Gnedin et al. 2001), 
can be accommodated.
There are many observations, like the HST observation giving 
$H_0=0.72\pm3$(stat)$\pm7$(systematic) km s$^{-1}$ Mpc$^{-1}$ 
(Freedman et al. 2001), which give a rather higher value of $H_0$.
However, there are several observations which also measure smaller 
values of $H_0$. For example, there is another HST Key Project which gives
$H_0=64^{+8}_{-6}$ km s$^{-1}$ Mpc$^{-1}$ (Jha et al. 1999).
Sandage and his collaborators find a value even as low as 
$H_0=58\pm6$ km s$^{-1}$ Mpc$^{-1}$ from an analysis
of SNe Ia distances (Parodi et al. 2000).

Additionally, unlike the $\Lambda$-dominated models, the EdS model has no 
strong 
integrated Sachs-Wolfe effect, so is in better agreement with the low 
quadrupole seen by WMAP (Blanchard, 2003).

\noindent
(ii) The recent data on
distant x-ray clusters obtained from XMM and Chandra projects
indicate that the observed
abundances of clusters at high redshift, taken at face value, 
give $0.9<\Omega_{\rm m0}<1.07$ (at 1 $\sigma$) (Blanchard, 2005). 
This is in striking agreement with a matter-dominated EdS model and 
is hard to reconcile with the concordance $\Lambda$CDM model.

\noindent
(iii) The estimated value of the parameter $\kappa \rho_{\rm g0}H_0^{-1}$ 
from the SNe fit is indeed of order of unity, as expected from
the theoretical reasoning described in section 2. 

\noindent
(iv)
Unlike the models of dark energy, the EdS model is not very speculative.

\bigskip
In order to have a visual comparison of the fits of different models to the
actual data points, we
magnify their differences by plotting the relative magnitude with respect to
a fiducial model ($\Omega_{\rm m0}=0$, $\Omega_{\Lambda0}=0$, without 
whiskers), [which has a reasonably good fit ($\chi^2$/dof $=191.701/156=1.2$) 
with ${\cal M}=43.40$]. This has been shown
in the `modified' Hubble diagram in Figure 1.
In Figure 2, we have shown the allowed regions in the parameter space 
$\kappa \rho_{\rm g0}H_0^{-1} - {\cal M}$ at 95\% and 99\% confidence
levels. 

\begin{figure}[tbh!]
\centerline{{\epsfxsize=14cm {\epsfbox[50 250 550 550]{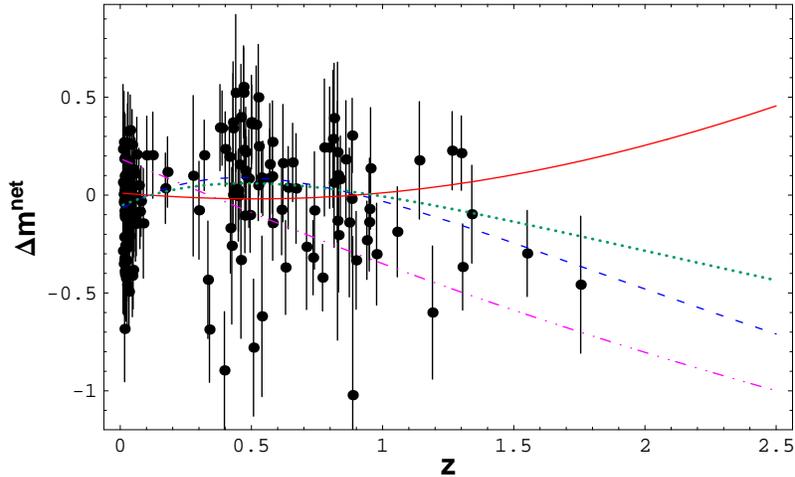}}}}
{\caption{\small Hubble diagram of the `gold sample' of SNe Ia minus a 
fiducial model ($\Omega_{\rm m0}=0$, $\Omega_{\Lambda0}=0$): 
The relative magnitude($\Delta m^{\rm net}\equiv m^{\rm net}-
m_{\rm fiducial}$) is plotted for some best-fitting models.
The solid curve corresponds to the EdS model with the whisker dust, the 
dotted curve corresponds to the flat $\Lambda$CDM model, the dashed curve 
corresponds to the spherical $\Lambda$CDM model, and the dashed-dotted
curve corresponds to the EdS model without any whisker dust. Some SNe
with redshift between 0 and 1 (mostly with $z\sim 0.5$) are missed by all the 
models and seem to be outliers. 
}}
 \end{figure}

\bigskip

\begin{figure}[tbh!]
\centerline{{\epsfxsize=14cm {\epsfbox[50 250 550 550]{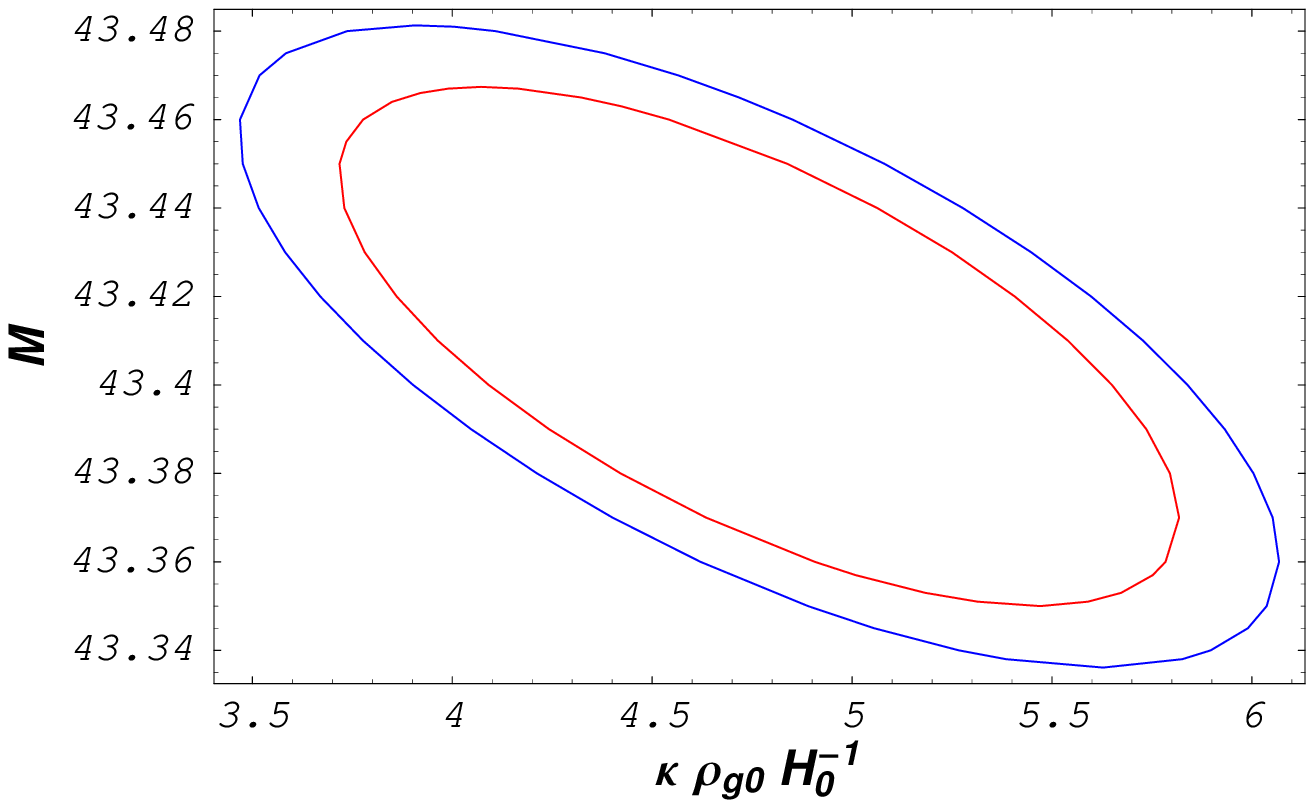}}}}
{\caption{\small The allowed regions by the `gold sample' of SNe Ia data 
(with 157 points) are shown in the parameter space 
$\kappa \rho_{\rm g0}H_0^{-1}-{\cal M}$ of the Einstein de Sitter model.
The parameters $\kappa$, $\rho_{\rm g0}$ and $H_0$ are 
measured in units of $10^5$ cm$^2$ g$^{-1}$, $10^{-34}$ g cm$^{-3}$ and
100 km s$^{-1}$ Mpc$^{-1}$ respectively. The inner ellipse denotes
the 95\% confidence region and the outer one, the 99\% confidence region.}}
 \end{figure}

\bigskip
\noindent
{\bf 4. BAYESIAN APPROACH}

\noindent
The frequentist's goodness-of-fit test of models, described in the previous
section, uses the best-fitting parameter values to evaluate relative merits
of the models under consideration. Thus it judges only the maximum likely 
performance of the models.
The Bayesian theory, on the contrary, does not hinge upon the  best-fitting
parameter values and evaluates the overall performance of the models. It has 
thus appeared as a powerful tool of model comparison.
The \emph{Bayes factor}, which is a  ratio of average likelihoods (rather
than the maximum likelihoods used for model comparison in frequentist 
statistics) is given by
\be
B_{ij}=\frac{{\cal L}(M_i)}{{\cal L}(M_j)} \equiv\frac{p(D|M_i)}{p(D|M_j)},
\label{eq:BF}
\ee 
where the likelihood for the model $M_i$, ${\cal L}(M_i)$ is the probability
$p(D|M_i)$ to obtain the data $D$ if the model $M_i$ is the true one (for
more details on the Bayesian theory, see Drell et al, 2000; John and 
Narlikar, 2002). For a model $M_i$ with free parameter, say, $\alpha$ and
$\beta$ (generalization for the models with more parameters is straight 
forward), this probability is given by 
\be
{\cal L}(M_i) \equiv p(D|M_i)=\int {\rm d}\alpha \int {\rm d}\beta~~ 
p(\alpha|M_i)~p(\beta|M_i)~ {\cal L}_i(\alpha, \beta),\label{eq:LP}
\ee 
where $p(\alpha|M_i)$ and $p(\beta|M_i)$ are the prior probabilities for the 
parameters $\alpha$ and $\beta$ respectively, 
assuming that the model $M_i$ is true. ${\cal L}_i(\alpha, \beta)$ is the 
likelihood for $\alpha$ and $\beta$  in the model $M_i$ and is usually given 
by the $\chi^2$-statistic:
\be
{\cal L}_i(\alpha, \beta)=\exp\left[-~\frac{\chi^2_i(\alpha, \beta)}{2}\right],
\ee 
where $\chi^2_i$ is the $\chi^2$-statistic for the model $M_i$, like the one
for the EdS model given by equation (\ref{eq:chi}). For flat prior 
probabilities for the parameters $\alpha$ and $\beta$, i.e., assuming that we 
have no prior information regarding $\alpha$ and $\beta$ except that they lie 
in some range [$\alpha$, $\alpha+\Delta\alpha$] and [$\beta$,
 $\beta+\Delta\beta$], we have $p(\alpha|M_i)=1/\Delta\alpha$ and 
$p(\beta|M_i)=1/\Delta\beta$. Hence the expression for the likelihood of the 
model $M_i$ reduces to
\be
{\cal L}(M_i) =\frac{1}{\Delta\alpha}\frac{1}{\Delta\beta}\int_\alpha^{\alpha+
\Delta\alpha} \int_\beta^{\beta+\Delta\beta} \exp\left[- ~\frac{\chi^2_i(
\alpha,\beta)}{2}\right]~{\rm d}\beta ~ ~{\rm d}\alpha .\label{eq:likeli}
\ee
The \emph{Bayes factor} $B_{ij}$, given by (\ref{eq:BF}), which measures the 
relative merits of model $M_i$ over model $M_j$, is interpreted as follows. If
$1<B_{ij} <3$, there is an evidence against $M_j$ when compared with 
$M_i$, but it is not worth more than a bare mention.  If $3<B_{ij} <20$, the
evidence   against $M_j$ is definite but not strong.  For $20<B_{ij} <150$, 
this evidence is strong and 
 for $B_{ij}>150$, it is very strong.

In order to compare the $\Lambda\neq0$ cosmology with  EdS one, we presume 
that the universe is flat spatially. This gives a fair comparison as well, 
since both the 
models are similar and have the same number (two) of free parameters.
As we do not have any prior information on the common parameter ${\cal M}$,
we take help from it's best-fitting values, which lie between 43 and 44.
Hence we assign a flat prior on ${\cal M}$ that it lies in the range 
[41, 45]. (This probability does not have any consequences though, as it 
cancels out in the Bayes factor.) Similarly, for the other parameter 
$\kappa \rho_{\rm g0}H_0^{-1}$
in the EdS model (the best-fitting value $\sim 4.8$), we assume that
$\kappa \rho_{\rm g0}H_0^{-1}\in [0, 10]$. For the parameter $\Omega_{\rm m0}$
in the flat $\Lambda$CDM model (the best-fitting value $\sim 0.3$), we
assume that $\Omega_{\rm m0}\in$ [0, 1]. When calculated for the `gold sample',
this gives a Bayes factor favouring
a model with $\Lambda\neq0$ over one with $\Lambda=0$ (EdS) as $B=1.98$, 
which though
indicates an evidence against the EdS model, but not more than a bare mention.

However, if no prior assumptions are made about the spatial geometry of the 
universe, there is a strong evidence against the EdS model: In this case, we
assign the prior probabilities $\Omega_{\rm m0}\in$ [0, 3], 
 $\Omega_{\Lambda0}\in [-3, 3]$, as have also been considered by Drell et al
(2000) and John \& Narlikar (2002). This is because the best-fitting values 
are towards higher side. This gives a Bayes factor for $\Lambda$CDM model
compared to EdS model as $B=113.23$, for a 
conservative prior probability $\kappa \rho_{\rm g0}H_0^{-1}\in [4, 5]$. This
shows a {\it strong} evidence against the EdS model, though it is not 
{\it very strong}. For a more liberal prior, however, this evidence moves
towards {\it very strong} side.

One should also note that a proper assessment of the probability for a model
is given by $p=1/(1+B)$. Thus if the universe is presumed to be flat, the
probability for the EdS model is 0.34, a reasonably good probability. 
In the absence of any prior 
assumption about the spatial geometry of the universe, this probability is
only 0.009 (which is still larger than 0.008 obtained from the frequentist 
approach). These results, together with the results and discussion of the 
previous section, make a reasonable case for the EdS model.

\bigskip
\noindent
{\bf 5. EFFECTS OF WEAK LENSING}

\noindent
Weak gravitational lensing is an unavoidable systematic uncertainty in the use
of SNe Ia as standard candles. As the universe is inhomogeneous in matter
distribution, the SNe fluxes are magnified by foreground galaxy excess and
demagnified by foreground galaxy deficit, compared to a smooth matter 
distribution. Recently Williams \& Song (2004) have reported such a correlation
between the magnitudes of 55 SNe from the sample of Tonry et al (2003) and
foreground galaxy overdensity. They have found the difference between the
most magnified and the most demagnified SNe as about 0.3-0.4 mag. Wang (2004)
has claimed further evidence of weak lensing in the high redshift sample of
SNe Ia from the Riess et al (2004) data. She has found a high magnification 
tail
at the bright end of the distribution, and a demagnification shift of the peak
of the distribution towards the faint end, which are signatures of weak 
lensing.
She has estimated the possible magnification of the three intrinsically most
luminous SNe (in the absence of lensing) in the bright end tail as:

\medskip
\noindent
SN1997as ($z=0.508$, $\mu_o=41.64$): magnified by $2.10\pm0.68$;\\
SN2000eg ($z=0.540$, $\mu_o=41.96$): magnified by $1.80\pm0.70$;\\
SN1998I  ($z=0.886$, $\mu_o=42.91$): magnified by $2.42\pm1.98$.

\medskip
\noindent
These high magnification factors $\sim 2$ from weak lensing are though somewhat
surprising, as also mentioned by Menard \& Dalal (2004) who claim not to find
any significant correlation between SN magnification and foreground galaxy 
overdensity. However, excluding these three lensed SNe from the sample
indeed improves the fit: $\chi^2$/dof = 192.49/152 = 1.27, $Q=0.015$ obtained 
for ${\cal M}=
43.41$ and $\kappa \rho_{\rm g0}H_0^{-1}=4.82$. The fit to the
$\Lambda$CDM models also improves considerably: 

\noindent
$\chi^2$/dof $= 164.20/151=1.09$, $Q=0.219$ obtained for $\Omega_{\rm m0}=0.47$, $\Omega_{\Lambda0}=1.02$, with ${\cal M}=43.32$;\\
$\chi^2$/dof $= 166.86/152=1.10$, $Q=0.194$ obtained for $\Omega_{\rm m0}=1-\Omega_{\Lambda0}=0.30$, with ${\cal M}=43.34$.

Though the weak lensing effects are estimated to be small for SNe at $z<1$,
they are non-negligible for higher redshift SNe. 
As more SNe are discovered at higher redshifts, it becomes increasingly
important to minimize the effect of weak lensing by, for example, using 
flux-averaging (Wang, 2000). 

\bigskip
\noindent
{\bf 6. CONCLUDING REMARKS}

\noindent
Since the early 1980s, inflation and cold dark matter have been the
dominant theoretical ideas in cosmology. However a key prediction of 
these ideas - the canonical Einstein-de Sitter model ($\Omega_{\rm m0}=1$, 
$\Lambda=0$)- seems to be in trouble in explaining the high redshift SNe
observations, as is widely believed. However, it can still explain these 
observations
successfully if one takes account of the extinction of light by intergalactic
metallic dust ejected from the SNe explosions, as has been shown earlier
(Vishwakarma, 2003) by considering Perlmutter et al' data (Perlmutter et al., 
1999). We have shown, in this paper, that the model also has an acceptable
fit to the recently published `gold sample' of 157 SNe Ia by Riess et al 
(2004) which, in addition to having previously observed SNe, also includes 
16 newly discovered SNe Ia, 6 of them being among the 7 highest redshift SNe 
Ia known so far, all at redshift $>1.25$. Though the fit to this new data
is deteriorated considerably, we believe that one should wait for
more accurate SNe Ia data with $z$ significantly $>1$ (as expected from
the coming space missions such as SNAP and JWST) in order to rule out a 
model which 
seems to explain all other existing observations well, some even better than 
the $\Lambda$CDM models. This point of view is also corroborated by
the Bayesian analysis of the data, which indicate that there is no significant
evidence against the EdS model if one assumes a spatially flat universe.
Even in the absence of such an assumption, there is not a very 
strong evidence against this model, for suitable priors.

We should also examine more critically, than has been done hitherto, the 
assumption
of a non-evolving standard candle for SNe Ia with high redshifts.
Though most studies confirm that the luminosity properties of SNe at different
redshift and environments are similar (Perlmutter et al., 1999; Sullivan et 
al, 2003), however, there are other theoretical studies which have found 
variations indicating evolutionary effects (Dominguez et al, 1998; Hoflich et 
al, 1998).
Additionally, it has been shown by Drell et al (2000) that a comparison of 
the peak 
luminosities estimated for individual SNe Ia by two different methods
are not entirely consistent with one another at high redshifts, $z\sim 0.5$. 
If evolution was entirely absent, the differences between them
should not depend on redshift. Moreover the three luminosity estimators 
in practice (the multicolor light curve shape method, the template fitting
method, and the stretch factor method) reduce the dispersion of distance 
moduli about best fit models at low redshift, but they do not at high 
redshift, indicating that the SNe may have evolved with redshift
(Drell et al, 2000). 

    The possible role of gravitational lensing in amplifying the 
supernova luminosity at high redshifts has been discussed by several authors 
and we have applied those ideas here to illustrate the difference it can make 
to any conclusion drawn from the data.  
Additionally, the role of 
intergalactic whisker dust still remains to be appreciated fully and we have 
demonstrated here the possible difference it can make to the viability of 
a model.

One may think that the whisker dust can create too much optical depth 
for the high redshift objects and they need to be excessively bright in
order to be seen. However, from our calculations, we find that even the 
objects with redshift as high as 5 will be fainter by $\sim$4 magnitudes only,
with the kind of whisker dust (best-fitting value) we have been talking about.
Thus the microwave emission from the high redshift quasars will also be
visible without unrealistically high demands on their luminosities.

\bigskip
\noindent
{\bf ACKNOWLEDGEMENTS}

\noindent 
The author thanks the Abdus Salam ICTP for sending 
the necessary literature under author's associateship programme. Thanks are 
also due to Moncy V. John for discussion, to Victor M. Ba$\tilde{n}$uelos 
Alvarez for 
providing help in computation, and to an anonymous referee for useful
comments.

\bigskip
\noindent
{\bf REFERENCES}\\
Aguire A. N., 1999, ApJ, 512, L19\\
Banerjee S. K., Narlikar J. V., Wickramasinghe N. C., Hoyle F., Burbidge,

\hspace{.5cm} G., 2000, ApJ, 119, 2583

\noindent
Blanchard A., Douspis M., Rowan-Robinson M., Sarkar S., 2003, A\&A, 412, 

\hspace{.5cm} 35

\noindent
Blanchard A.,2005, preprint, astro-ph/0502220\\
Cayrel R., et al, 2001, Nature, 409, 691\\
Chitre, S. M., Narlikar J. V., 1976, Astrophys. Space Sc., 44, 101\\
Csaki C., Kaloper N., Terning J., 2002, Phys. Lett. B 535, 33\\
Csaki C., Kaloper N., Terning J., 2004, preprint, astro-ph/0409596\\
Dominguez I., et al, 1998, preprint, astro-ph/9809292\\
Donn B., Sears G. W., 1963, Science, 140, 1208\\
Drell P. S., Loredo T. J., Wasserman I., 2000, ApJ, 530, 593\\
Freedman W. L. et al., 2001, ApJ., 553, 47\\
Gnedin O. Y., Lahav O., Rees M. J., astro-ph/0108034\\
Hoflich P., Wheeler J. C., Thielemann F. K., 1998, ApJ, 495, 617\\
Hoyle F., Wickramasinghe N. C., 1988, Astrophys. Space Sc. 147, 245\\
Hoyle F., Burbidge G., Narlikar J. V., 2000, {\it A Different Approach to

\hspace{.5cm}       Cosmology}, (Cambridge: Cambridge Univ. Press)

\noindent
Jha S., et al., 1999, ApJ. Suppl., 125, 73\\
John M. V., Narlikar J. V., 2002,  Phys. Rev. D, 65, 043506\\
Menard B., Dalal N., 2004, preprint, astro-ph/0407023\\
Nabarro F. R. N., Jackson P. J., 1958, in \emph{Growth and Perfection in Crystals}, 

\hspace{.5cm}  eds. R. H. Duramus, et al, (J. Wiley, New York)

\noindent
Narlikar J. V., Wickramasinghe N. C., Sachs R., Hoyle F., 1997, Int. J. Mod.

\hspace{.5cm}     Phys. D, 6, 125

\noindent
Narlikar J. V., Vishwakarma, R. G., Burbidge G., 2002, PASP, 114, 1092\\
Parodi B. R., et al., 2000, ApJ., 540, 634\\
Perlmutter S., et al., 1999, ApJ., 517, 565\\
Press W. H., Teukolsky S. A., Vetterling W. T., Flannery B. P., 1986, 

\hspace{.5cm} {\it Numerical Recipes}, (Cambridge University Press) 

\noindent
Riess A. G., et al., 2004, ApJ., 607, 665\\
Sullivan M., et al, 2003, MNRAS, 340, 1057\\
Tonry J. L., et al, 2003, preprint, astro-ph/0305008\\
Vishwakarma R. G.,  2002, MNRAS, 331, 776\\
Vishwakarma R. G., 2003, MNRAS, 345, 545\\
Wang Y., 2000, ApJ., 536, 531\\
Wang Y., 2004, preprint, astro-ph/0406635\\
Wickramasinghe N. C., Wallis D. H., 1996, Astrophys. Space Sc.

\hspace{.5cm} 240, 157

\noindent
Williams L. L. R., Song J., 2004, preprint, astro-ph/0403680\\
\end{document}